\def\year{2020}\relax
\documentclass[letterpaper]{article} 
\usepackage{aaai20}  
\usepackage{times}  
\usepackage{stackengine}

\usepackage{tabularx}
\usepackage{helvet}  
\usepackage{courier}  
\usepackage{url}  
\usepackage[final]{graphicx}  
\usepackage{array,multirow}
\usepackage{xcolor}
\usepackage{array}
\usepackage{booktabs}
\usepackage{enumitem,kantlipsum}
\usepackage[normalem]{ulem}
\usepackage{hhline}
\usepackage{mathtools}
\usepackage{xspace} 
\usepackage{} 
\usepackage{natbib}
\usepackage{color, soul}
\usepackage{subfigure}

\usepackage[textsize=scriptsize,backgroundcolor=green]{todonotes}
\newcolumntype{L}[1]{>{\raggedright\let\newline\\\arraybackslash\hspace{0pt}}m{#1}}
\newcolumntype{C}[1]{>{\centering\let\newline\\\arraybackslash\hspace{0pt}}m{#1}}
\newcolumntype{R}[1]{>{\raggedleft\let\newline\\\arraybackslash\hspace{0pt}}m{#1}}
\begin{document}
%

\title{News Consumption in Time of Conflict: 2021 Palestinian-Israel War as an Example}

\author{
Kareem Darwish
\\
    QCRI, HBKU, Doha, Qatar \\
    kdarwish@hbku.edu.qa
}



\maketitle
\begin{abstract}
This paper examines news consumption in response to a major polarizing event, and we use the May 2021 Israeli-Palestinian conflict as an example.  We conduct a detailed analysis of the news consumption of more than eight thousand Twitter users who are either pro-Palestinian or pro-Israeli and authored more than 29 million tweets between January 1 and August 17, 2021. We identified the stance of users using unsupervised stance detection. We observe that users may consume more topically-related content from foreign and less popular sources, because, unlike popular sources, they may reaffirm their views, offer more extreme, hyper-partisan, or sensational content, or provide more in depth coverage of the event. The sudden popularity of such sources may not translate to longer-term or general popularity on other topics.

\end{abstract}

\section{Introduction}
The emergence of a major polarizing event attracts the attention of social media users resulting in a flurry of interactions such as posts, likes, retweets, and shares.  Such interactions often stem from the desire of users to express their stances and opinions on the polarizing event.  One of the important forms of expression is content sharing from different media sources that support the views of users.  Thus, users may turn to sympathetic media sources that provide greater coverage, in the form of news articles, opinion pieces, video clips, documentaries, or podcasts. Depending on the event of interest, the media sources may be well-established, foreign, or less popular.  In this paper, we examine the reaction of US Twitter users to the conflagration of hostilities between Israelis and Palestinians in May 2021, and we focus primarily on users' sharing activities from media sources that are either foreign or generally less popular among US users.  This event has two distinctive characteristics, namely: 1) it is related to a foreign issue that may not receive continuous coverage in popular news sources, but is yet highly polarizing; and 2) popular US media sources are generally sympathetic to the Israeli narrative, including left-leaning media such as CNN\footnote{\url{https://mediabiasfactcheck.com/left/cnn-bias/}} \cite{ezzina2021western}, while many progressive voices are sympathetic to the Palestinian narrative \footnote{\url{https://aje.io/lep74}}.  These two characteristics may cause users to express clear stances by sharing content from media sources that reaffirm their views, including foreign or less popular.  In the context of the May 2021 Israeli-Palestinian conflict, we are interested in answering the following questions:
\begin{enumerate}
    \item With the advent of a major polarizing events, are social media users likely to share content from foreign or less popular media sources?
    \item If users are likely to do so, what is the reason? Is it due to interest in deeper coverage or mismatch in stance alignment with popular sources?
    \item Does the sudden popularity of the sources translate into future longer-term popularity?
\end{enumerate}
To answer these questions, we collected 6.9 million tweets pertaining to the May 2021 Israeli-Palestinian conflict, and we used stance detection to identify the stances of the most vocal 8,034 users as either pro-Palestinian or pro-Israeli.  We then proceeded to acquire the historical tweets of these users that cover the period from January 1, 2021 to August 17, 2021 to measure their relative consumption of different right and left-leaning sources of varying popularity before, during, and after the conflict.

Our contributions in this work are as follows:
\begin{itemize}
    \item We analyze the news consumption of thousands of users over a span of several months in reference to a major polarizing event.  This includes automatically identifying users with opposing stances along with the frames that they respond to in the news.
    \item We show that users may resort to foreign or less popular, but potentially more focused, media, and we attempt to explain the motivations of users based on their interactions with different media sources.
    \item We show that the sudden popularity of some sources on specific topics may not translate to longer-term or broader popularity on other topics, and we explore possible reasons for this.
\end{itemize}

\section{Background}
\paragraph{News Popularity}
Much of the work on news popularity has focused on predicting future popularity of individual news items \citep{bandari2012pulse,hensinger2013modelling,keneshloo2016predicting}. Prior work has shown that the frequency of interactions among Twitter users with a news item is correlated with the popularity of that item \citep{wu2015analyzing} and can be used to predict the size of the items's readership (or viewership) \citep{castillo2014characterizing}. Thus, social media popularity is indicative of real-world popularity. As for the popularity of news media sources, as opposed to individual items, prior worked has focused on a variety of aspects such as: the relative popularity of established versus upcoming media organizations, where older well-established organization have an advantage over newer less-recognized ones \citep{nelson2020enduring}; the emergence of online news media, where print media has been in decline for years \citep{chyi2019analog}; the polarizing effect of social media in increasing the popularity of more extreme sources (and views) \citep{warner2014echoes}; whether users consume news from ideologically homogeneous or heterogeneous sources \citep{mullainathan2005market,nelson2017myth}; and the correlation between the ideological positions or stances of users and the news they share, where such correlation was used to estimate the political leaning of news sources \citep{stefanov2020predicting}.  In this work, we focus the effect of polarizing events on the relative popularity of news sources.

\paragraph{Stance Detection} 
Stance detection can be performed 
using supervised classification with a variety of features, such as text-level features (e.g., words or hashtags), user-interaction features (e.g., user mentions and retweets), and profile-level features (e.g., name and location) \citep{borge2015content,magdy2016isisisnotislam,magdy2016failedrevolutions}. 
The use of retweets seems to yield competitive results \citep{magdy2016isisisnotislam,wong2013quantifying,wong2016quantifying}. 
Label propagation is also an effective semi-supervised method that propagates labels in a network based on follow or retweet relationships \citep{borge2015content,weber2013secular} or the sharing of identical tweets \citep{darwish2018scotus,kutlu2018devam,magdy2016isisisnotislam}. 
More recent work projects user onto a two dimensional space and then uses clustering to perform unsupervised stance detection \citep{darwish2019unsupervisedStance}, with the best setup involving the use of UMAP for projection, mean shift for clustering, and the retweeted accounts as user features. Though this method may have relatively low recall, it has nearly perfect precision.  In this paper, we use the published implementation of this method\footnote{\url{https://github.com/p-stefanov/stance-detection}} to automatically label users. Detecting the stance of Twitter users has been shown to be effective in ascertaining the leaning of media sources overall and on specific topics, as users tend to cite sources that agree and/or reaffirm their beliefs \citep{stefanov2020predicting}.

\paragraph{May 2021 Israeli-Palestinian Conflict} In April 2021, tensions between Palestinians and Israelis increased due to two main factors. The first is related to the threatened expulsion of 75 Palestinian families from the Sheikh Jarrah neighborhood in Jerusalem.  The issue was framed by the Israeli authorities as a property dispute between Palestinians and Israelis to be settled through court proceedings, while the Palestinians dubbed it as ethnic cleansing of Jerusalem that is carried out by an apartheid system. The second is related to repeated clashes between Palestinian worshippers and Israeli forces in and around Al-Aqsa Mosque Compound, which is one of the holiest sites for Muslims. The clashes started in April 2021 with Israeli efforts to partially block entry into Al-Aqsa Mosque and to limit the number of worshippers and culminated with Israeli police repeatedly storming Al-Aqsa compound on May 7, 8, and 10. Hamas, the ruling authority in the Gaza Strip, delivered an ultimatum to Israel to cease the expulsions in Sheikh Jarrah and to stop incursions into Al-Aqsa Compound by May 10.  With the Israeli government ignoring the ultimatum, Hamas along with other Palestinian factions in Gaza launched hundreds of rockets into Israel, sparking a full blown war between Palestinians and Israelis that lasted until a ceasefire agreement was reached on May 21.

\section{Dataset}
We collected tweets on May 19, 2021 related to the Isreali-Palestinian conflict using the Twitter search API using the following keywords: Palestine, Palestinian(s), Gaza, Israel, Zionist, Quds (Arabic name for Jerusalem), Alquds, Jerusalem, Netanyahu,  Israeli(s), and Jarrah along with their equivalents in Arabic. In all we collected 6,944,076 tweets as follows:
\begin{center}
\begin{tabular}{c|r}
    Date & Count \\ \hline
    May 17 & 537,571 \\
    May 18 & 3,822,528 \\
    May 19 & 2,576,534 \\
\end{tabular}
\end{center}

\begin{figure}[ht]
    \centering
        \begin{tabular}{c}
    \includegraphics[width=0.9\linewidth]{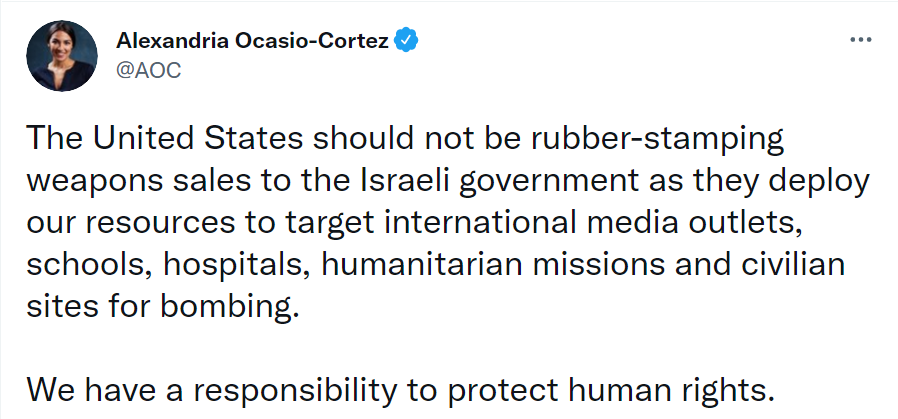} \\
    \vspace{6pt} \\
    \includegraphics[width=0.9\linewidth]{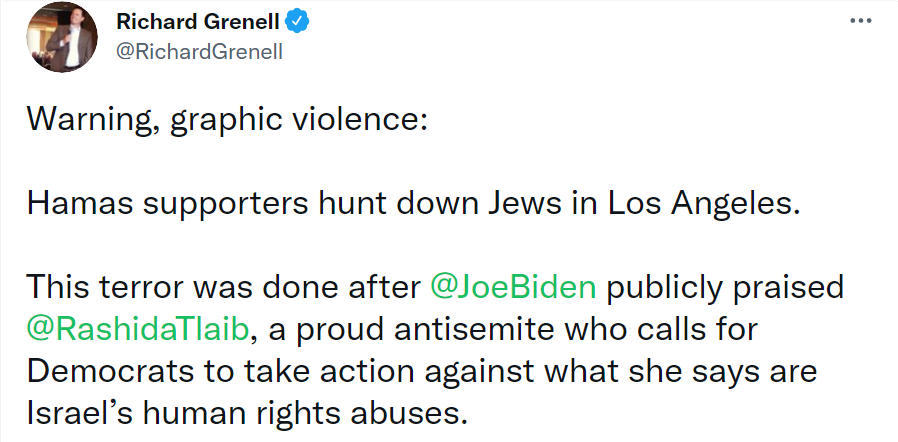} \\
        \end{tabular}
    \caption{Sample tweets for pro-Palestinian and pro-Israeli clusters.}
    \label{fig:sampleTweets}
\end{figure}

Next, we filtered tweets to retain tweets that were authored in English by US users.  To filter by language, we used the language labels provided by Twitter.  As for location filtering, we examined the users' declared locations, where we considered a user from the US if his/her location matched a manually curated list of 16,383 US locations (ex. ``Pinellas, Florida, USA'' and ``Prev:NJ/VA/PT/CA,now:Lville,ky'') or matched the pattern ``city name, (state name$\mid$abbreviation)'' (ex. ``Chicago, Ill.'', ``Chicago, IL'', and ``Little Rock, Arkansas''). After filtration, we were left with 357,725 tweets from 130,110 different users.  Using unsupervised stance detection \cite{darwish2019unsupervisedStance} over the most active 20,000 users, we attempted to identify underlying groups with varying stances.  
The method was able to identify the stances of 8,034 users, with 6,495 siding with the Palestinian side and 1,539 siding with the Israeli side.  Figure \ref{fig:sampleTweets} show the most retweeted tweets for both groups respectively. Table \ref{tab:mostCitedSites} shows the 10 most cited news sources for both groups along with their ideological leaning and credibility as provided by \url{mediabiasfactcheck.com}.  What is noteworthy in the list of most cited sources is the prominence of: foreign news sources (ex. the Guardian (\url{theGuardian.com} -- UK), AlJazeera (\url{AlJazeera.com} -- Qatar), and Jerusalem Post (\url{JPost.com} -- Israel)); and less popular news sources that typically have limited reach (ex. Electronic Intifada (\url{electronicIntifada.net} -- Alexa rank: 272,401\footnote{\url{https://www.alexa.com/siteinfo/electronicintifada.net}}) and National Review (\url{NationalReview.com} -- 9,657\footnote{\url{https://www.alexa.com/siteinfo/nationalreview.com}})) compared to sites with much higher reach such as the Washington Post (\url{WashingtonPost.com}) and Fox News (\url{FoxNews.com}) that have Alexa ranks of 208\footnote{\url{https://www.alexa.com/siteinfo/washingtonpost.com}} and 251\footnote{\url{https://www.alexa.com/siteinfo/foxnews.com}} respectively.  Another noteworthy observation is that the media sources in Table \ref{tab:mostCitedSites} for pro-Palestinian users are overwhelmingly left-leaning with mixed to high credibility, while those for pro-Israeli users are right-leaning with mostly mixed to low credibility. 

\begin{table*}
    \centering
    \small
    \begin{tabular}{l|l|l|l|l|l}
\multicolumn{3}{c|}{Pro-Palestinian} & \multicolumn{3}{c}{Pro-Israeli} \\ 
Source	&	Credibility	&	Leaning	&	Source	&	Credibility	&	Leaning	\\ \hline
WashingtonPost.com	&	Mostly Factual	&	Left-center	&	FoxNews.com	&	Mixed	&	Right	\\
\textit{\textbf{theGuardian.com}}	&	Mixed	&	Left-center	&	\textit{\textbf{JPost.com}}	&	Mostly Factual	&	Right-center	\\
Reuters.com	&	Very High	&	Center	&	Breitbart.com	&	Mixed	&	Extreme right	\\
\textit{\textbf{AlJazeera.com}}	&	Mixed	&	Left-center	&	FreeBeacon.com	&	Mixed	&	Right/Extreme right	\\
JewishCurrents.org	&		&		&	NYPost.com	&	Mixed	&	Right-center	\\
\textit{\textbf{MiddleEastEye.net}}	&	Mostly Factual	&	Left-center	&	DjhjMedia.com	&	Low	&	Extreme right	\\
DemocracyNow.org	&	High	&	Left	&	OANN.com	&	Low	&	Extreme right	\\
ElectronicIntifada.net	&	Mostly Factual	&	Left	&	Hannity.com	&	Low	&	Extreme right	\\
MondoWeiss.net	&	Mixed	&	Extreme left	&	DailyCaller.com	&	Mixed	&	Right	\\
theOnion.com	&	Satire	&		&	NationalReview.com	&	Mostly Factual	&	Right	\\
    \end{tabular}
    \caption{Most cited news sources for pro-Palestinian and pro-Israeli users along with credibility and ideological leaning (from \url{mediabiasfactcheck.com}. Foreign sources are italicized and bolded.}
    \label{tab:mostCitedSites}
\end{table*}

\paragraph{\textbf{Timeline Data Crawling:}}\label{sec_timeline_crawling} 
On July 1, 2021 and again on August 21, 2021, we crawled the timelines of the 8,034 users, who were automatically tagged as pro-Palestinian or pro-Israeli, using the twarc Python library\footnote{\url{https://github.com/DocNow/twarc}}, which is a Twitter API wrapper. We collected on two separate days to crawl as many tweets as possible for the users. Twitter typically allows the crawling of the last 3,200 tweets for a user. Depending on how active each user is, 3,200 tweets can cover days, months, or years. Hence, the number of collected tweets per day decreases as we go back in time.  In all, we collected a little over 29 million tweets from January 1, 2021 to August 17, 2021.  We also split the tweets by group (pro-Palestinian or pro-Israeli) and by topic (Palestinian-Israeli conflict related or not), where we filtered tweets using the following keywords: Israel, Israeli, Palestine, Palestinian, Quds, Alquds, Jerusalem, Hamas, and Jarrah. The breakdown of tweets is as follows:

\begin{center}
    \begin{tabular}{r|r|r}
         & pro-Palestinian & pro-Israeli \\ \hline
        Topically related & 1,375,675 &  373,347 \\
        Non-topically related & 20,597,511 & 6,723,035 \\ 
    \end{tabular}
\end{center}

\begin{figure}[ht]
    \centering
    \includegraphics[width=0.85\linewidth]{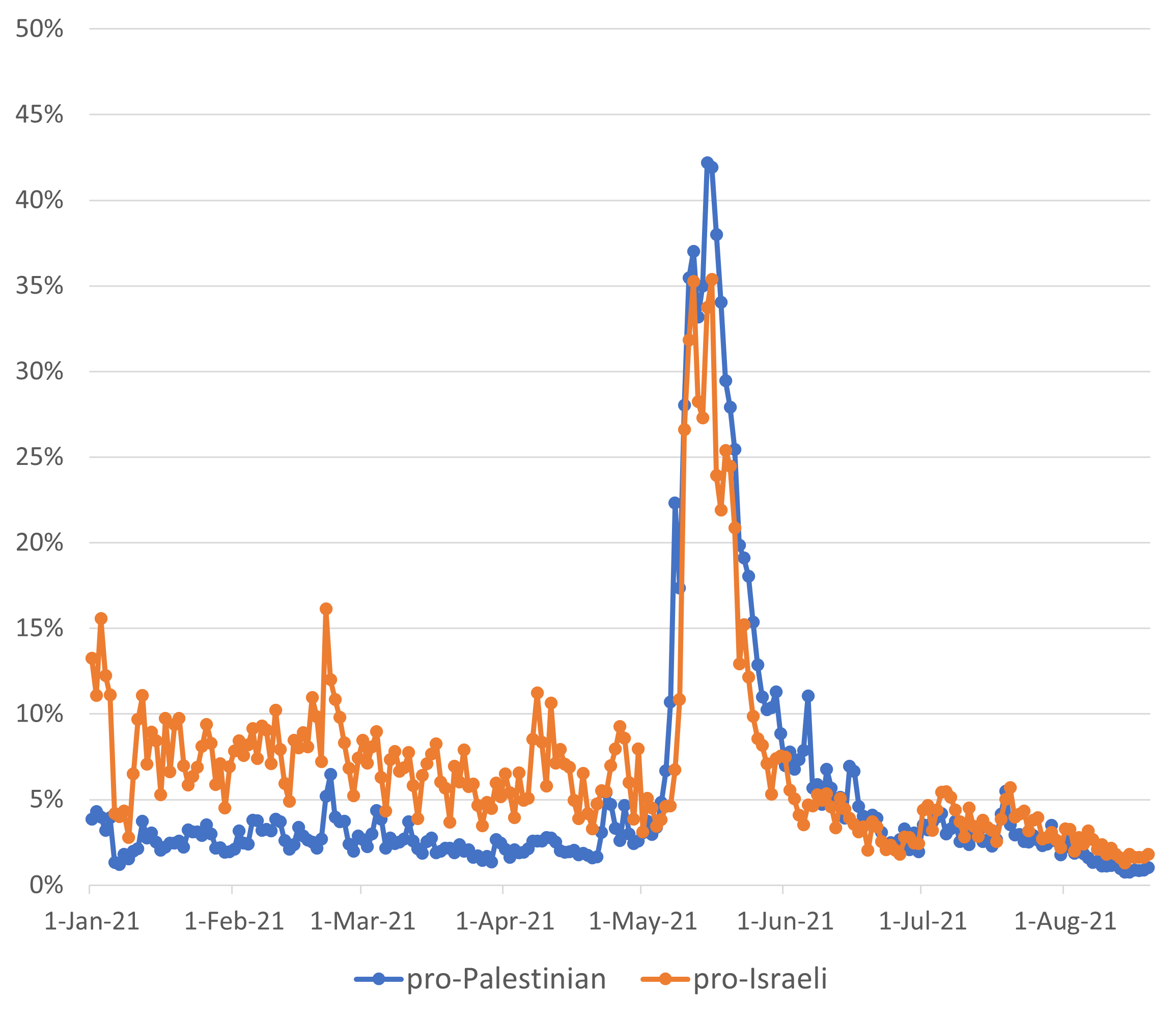}
    \caption{Percentage of topically related tweets for pro-Palestinian and pro-Israeli groups.}
    \label{fig:percentageOfTopicalTweets}
\end{figure}

Figure \ref{fig:percentageOfTopicalTweets} shows the percentage of topically related tweets over time for both groups.  It seems that the pro-Israeli group was generally more topically engaged than the pro-Palestinian from January to April, and the percentage of topically related tweets is evenly matched from May onward.  We checked the most frequent hashtags for the pro-Israeli group in the January to April time window, and 10 out of the top 30 hashtags were related to: the Holocaust (Auschwitz, YomHaShoah, and HolocaustRemembranceDay, and Holocaust); anti-Semitism (Antisemitism, JusticeForSarahHalimi, and SarahHalimi -- French Jewish doctor who was killed in a seemingly anti-Semitic attack); and general Judaism-related terms (ShabbatShalom, Jewish, and Passover).  Such  may indicate a large presence of Jewish users in the pro-Israeli group.  Other hashtags align with right-leaning or Republican talking points such as: Antifa, BidenBorderCrisis, WeveGotACountryToSave, GodBlessAmerica, GodBlessPresidentTrump, and ExpelMaxineWaters.
For the most frequent hashtags for the pro-Palestinian group, they were dominated by: progressive issues (MedicareForAll, ForceTheVote, FreeAssange, TransDayOfVisibility, CancelStudentDebt, AbolishThePolice, and RaiseTheWage) and foreign issues (FarmersProtest (India), Haiti, YemenCantWait, Somalia, FreePalestine, and SaveSheikhJarrah).

Going back to the central questions of the paper, the first question is: \textbf{Do users resort to foreign or less popular sources with the advent of a major polarizing event?}
To answer this question, we need to compare the relative citations of foreign and less popular sources in the periods prior, during, and post the event compared to popular sources.  For the pro-Palestinian group, we use the Washington Post (\url{WashingtonPost.com} -- Alexa rank: 208) and CNN (\url{cnn.com} -- 109\footnote{\url{https://www.alexa.com/siteinfo/cnn.com}}) as our reference popular media sources.  As for the pro-Israeli groups, we use Fox News (\url{FoxNews.com} -- 251) and Breitbart News (\url{breitbart.com} -- 415\footnote{\url{https://www.alexa.com/siteinfo/breitbart.com}}) as our reference. For foreign and less popular sources for the pro-Palestinian group, we use Aljazeera (\url{AlJazeera.com} -- Alexa rank: 1,659), Middle East Eye (\url{MiddleEastEye.net} -- 40,196), and Jewish Currents (\url{JewishCurrents.org} -- 292,953).  For the equivalent for the pro-Israeli group, we use Jerusalem Post (\url{JPost.com} -- Alexa rank: 4,487), Free Beacon (\url{FreeBeacon.com} -- 30,681), and Sean Hannity (\url{Hannity.com} -- 62,718).  We chose these media sources to cover foreign (non-US) and US sources with varying levels of popularity.  Users may cite news items in their tweets by either retweeting a source's tweets or by embedding a link to the site of the news source.  Many sources employ multiple Twitter accounts (ex. AlJazeera operates @AJEnglish, @AJENews, and @AJPlus) and use different base URLs (ex. Fox News uses \url{FoxNews.com} and \url{fxn.ws}).  We count a retweet or an embedded URL as a citation of a source.

\begin{figure*}[ht]
    \centering
    \includegraphics[width=.4\linewidth]{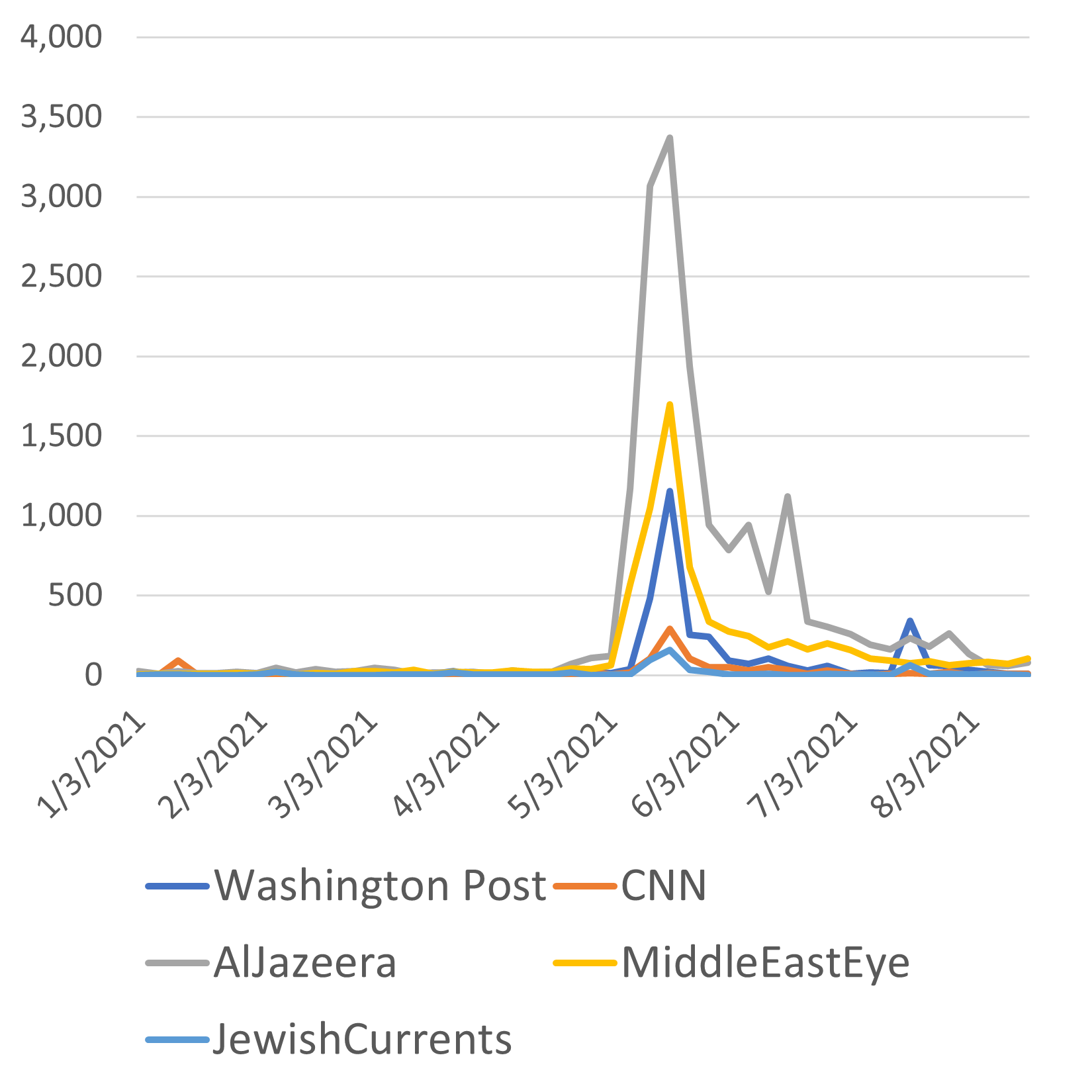} (a)
    \includegraphics[width=.4\linewidth]{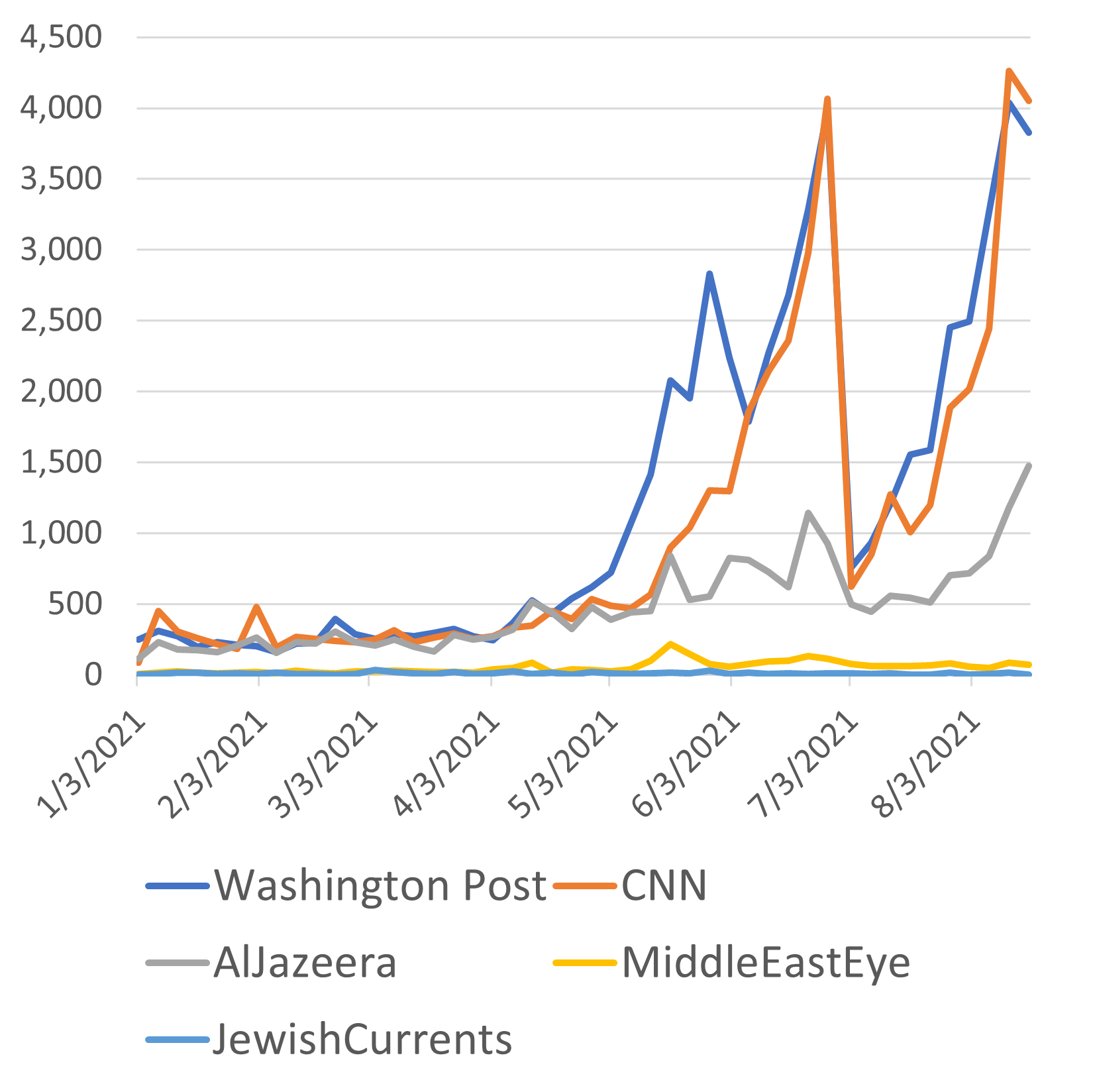} (b) \\
    \includegraphics[width=.4\linewidth]{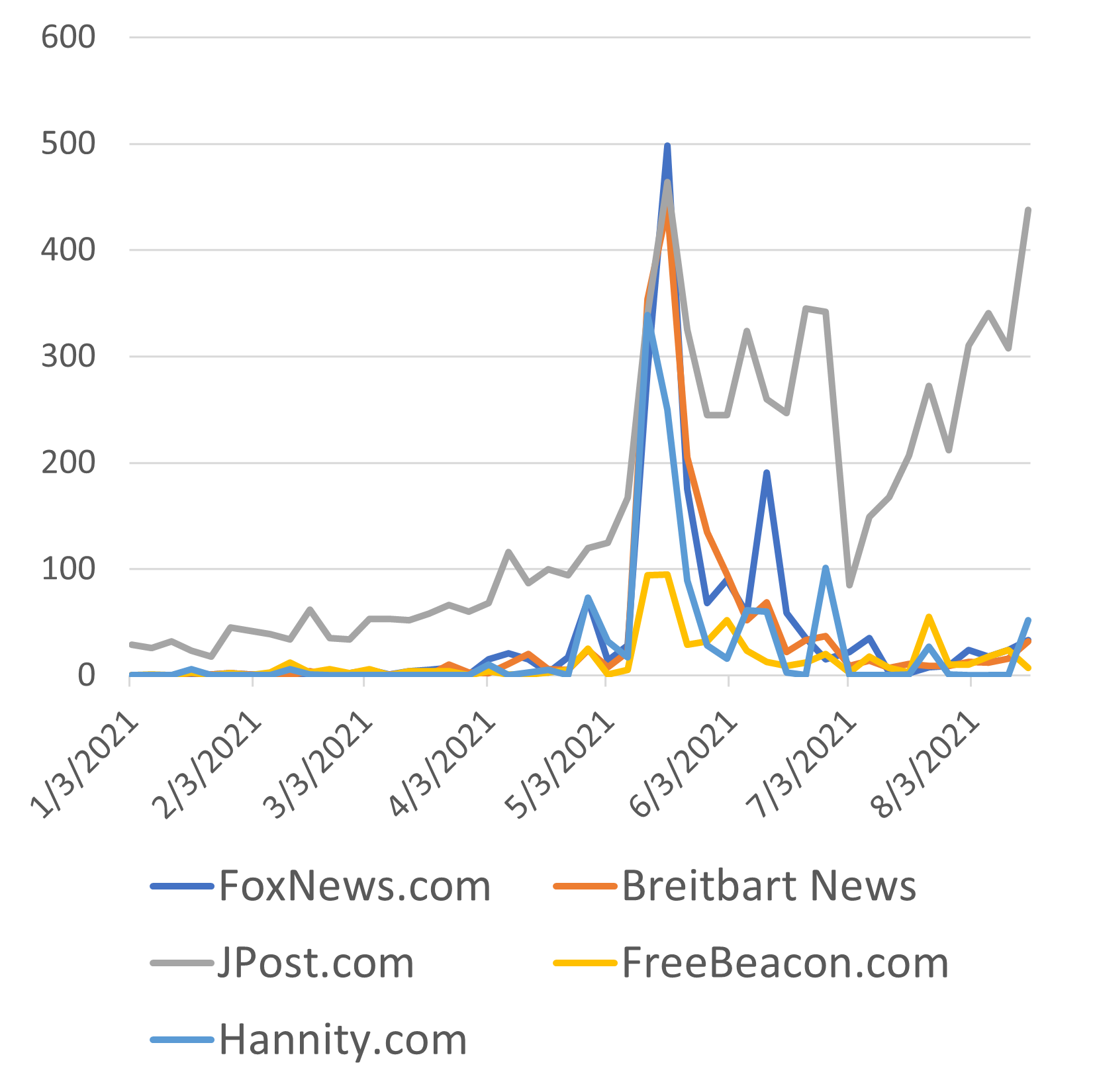} (c)
    \includegraphics[width=.4\linewidth]{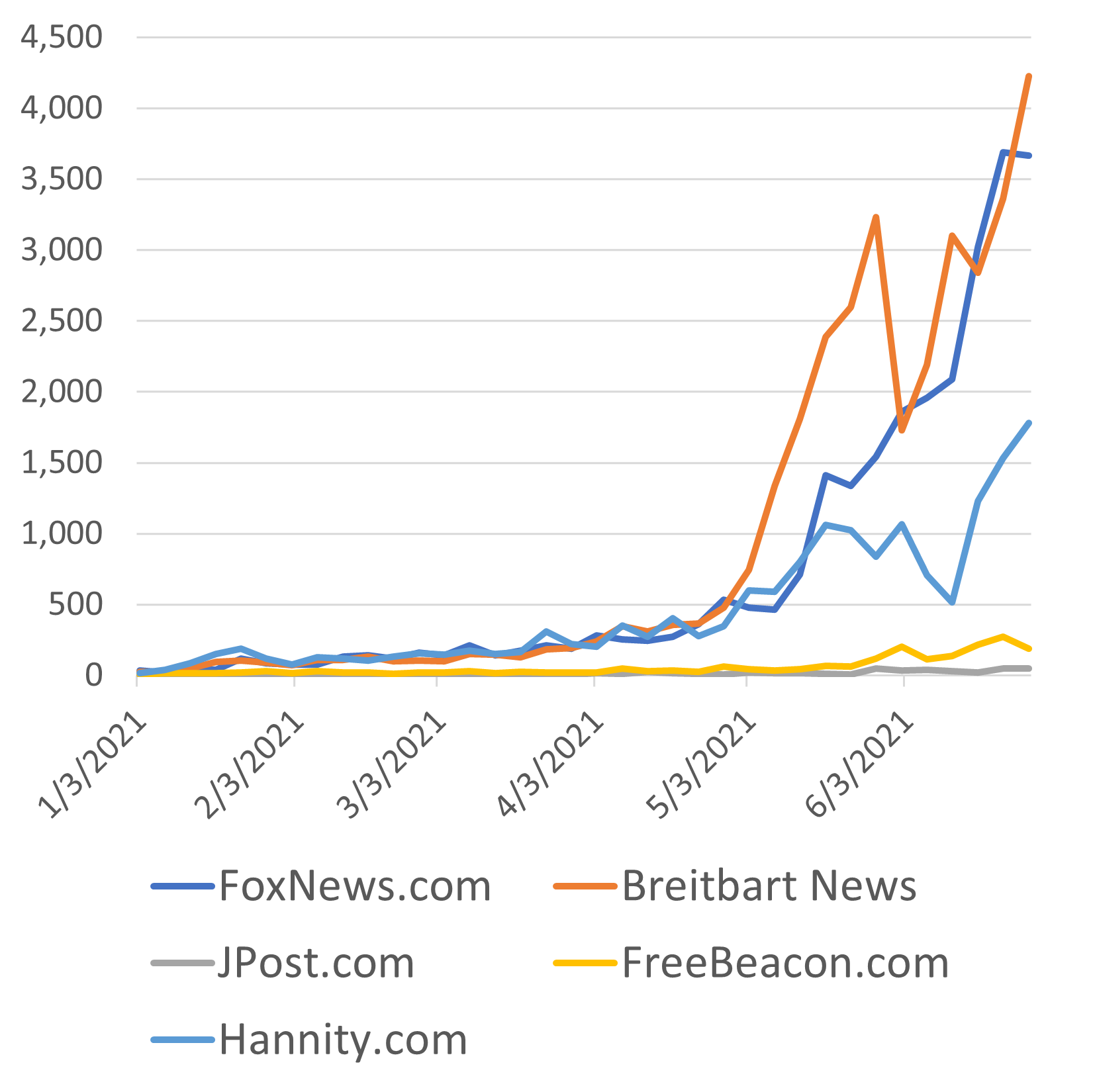} (d)
    \caption{Number of citations per source for pro-Palestinian group for (a) topic-related and (b) non-topic-related tweets, and for pro-Israeli group for (c) topic-related and (d) non-topic-related tweets. }
    \label{fig:citationsPerSource}
\end{figure*}

Figure \ref{fig:citationsPerSource} shows the number of citations for the aforementioned sources for pro-Palestinian and pro-Israeli groups with topically related and non-topically related tweets separated.  The sub-figures of the topically related citations show that the less popular and foreign sources match or eclipse the reference popular sources particularly during the period of the conflict and the ensuing period. This was more pronounced for well established foreign sources, namely AlJazeera and Jerusalem Post for the pro-Palestinian and pro-Israeli groups respectively.  Thus for our first question, the data suggests that users may indeed resort to foreign or less popular sources for their information needs particularly in conjunction with major polarizing events.

This leads us to our second question, namely: \textbf{Why do users tune into less popular or foreign sources? Is it because of depth of coverage or mismatch in stance alignment with popular sources?}  To answer this question, we examine topically related tweets that refer to the Israeli-Palestinian conflict during May 2021.  We look at all tweets that refer to one of the news sources of interest, whether in the form a retweet, reply, mention, or share of an article.  This would allow us to examine how pro-Palestinian and pro-Israeli users perceive the media sources.  For all media sources, we manually labeled the 25 most retweeted tweets.

\paragraph{Pro-Palestinian Group}
The Washington Post tweets can be roughly categorized into 4 categories, namely: 
\begin{itemize}
    \item Reporting news (45.6\%). Ex. RT @RashidaTlaib: No more weapons to kill children and families @JoeBiden. Enough. \url{https://t.co/QQcs2WYEhl} (article: Biden approves weapons sales to Israel).
    \item Exposing Israeli atrocities (27.9\%). Ex. RT @washingtonpost: The Committee to Project Journalists expressed concerns that Israel was "deliberately targeting media facilities in order to disrupt coverage of the human suffering in Gaza" \url{https://t.co/vgSkZ3z0fu} 
    \item Criticizing the Washington Post (21.0\%). Ex. RT @m7mdkurd: The @WashingtonPost podcast featuring me is highly edited. It’s amazing they used my quote saying “no one listens to Palestinians” \& then proceeded to remove “colonial” “fascist” “apartheid” from my vocabulary. They even took out the word “occupation.” 
    \item Praising coverage (5.5\%). Ex. RT @4noura: This is how it’s done. And this is how well get it done. @washingtonpost with an honest headline. \#GazaUnderAttack \#savesheikhjarah \#Jerusalem \#Haifa \#Lydd \#apartheidisrael \#Palestinian \#Freedom \url{https://t.co/haLMk8sXct} 
\end{itemize}
 
 As for CNN.com, tweets can be divided into 4 main categories, namely:
 \begin{itemize}
     \item Criticizing CNN (72.3\%). Ex. RT @Dena: This @CNN internal memo directs staff to say “Hamas-run Gaza Ministry of Health” when reporting casualty numbers. It was sent by the Jerusalem bureau chief.   This is a page straight out of Israel’s playbook. It serves to justify the attack on civilians \& medical facilities \url{https://t.co/nAWF4s7i1F}
     \item Referencing CNN staff (20.7\%).  Ex. RT @Stone\_SkyNews: The Israeli police pushing around accredited news correspondent @bencnn and his @cnni crew. It’s happened to us all this week.  Today I walked past a policeman.  I smiled and said hello. “F*ck off” he said. \url{https://t.co/FuS1D9g3ey} 
    \item Praising coverage (5.0\%). Ex. RT @4noura: I rarely get asked what it feels like to be Palestinian. Thank you @BeckyCNN \url{https://t.co/tHpWDaq9p8} 
    \item Reporting news (2.0\%).  Ex. RT @ryanobles: FIRST ON @CNN: Sen. Jon @ossoff leads a group of 28 members of the Democratic Senate Caucus calling for a cease fire in Israeli- Palestinian conflict. Full list of signatories: (W/ @jessicadean \& @DaniellaMicaela)
 \end{itemize}

\begin{figure*}[ht]
    \centering
    \includegraphics[width=.4\linewidth]{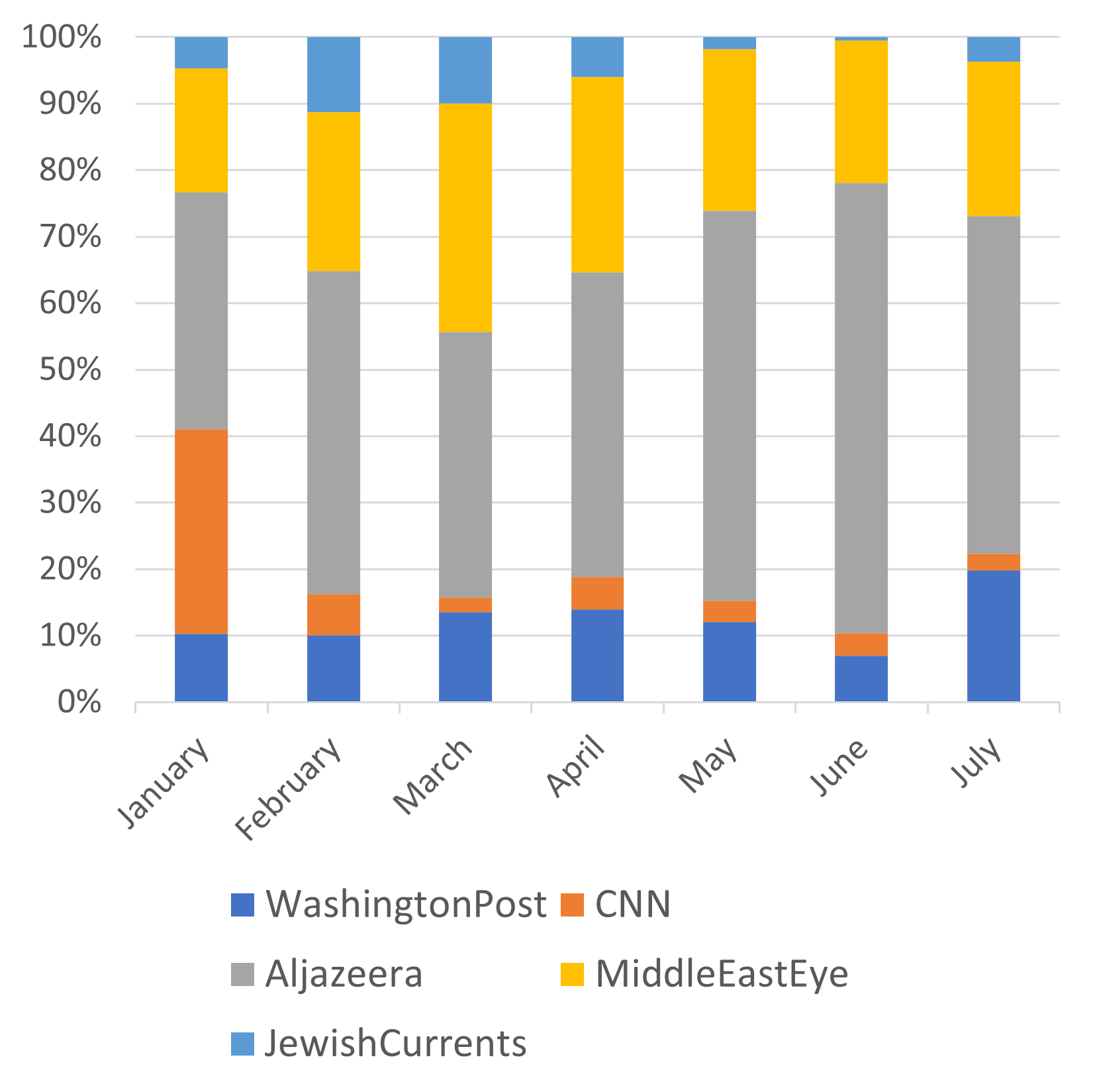}(a)
    \includegraphics[width=.4\linewidth]{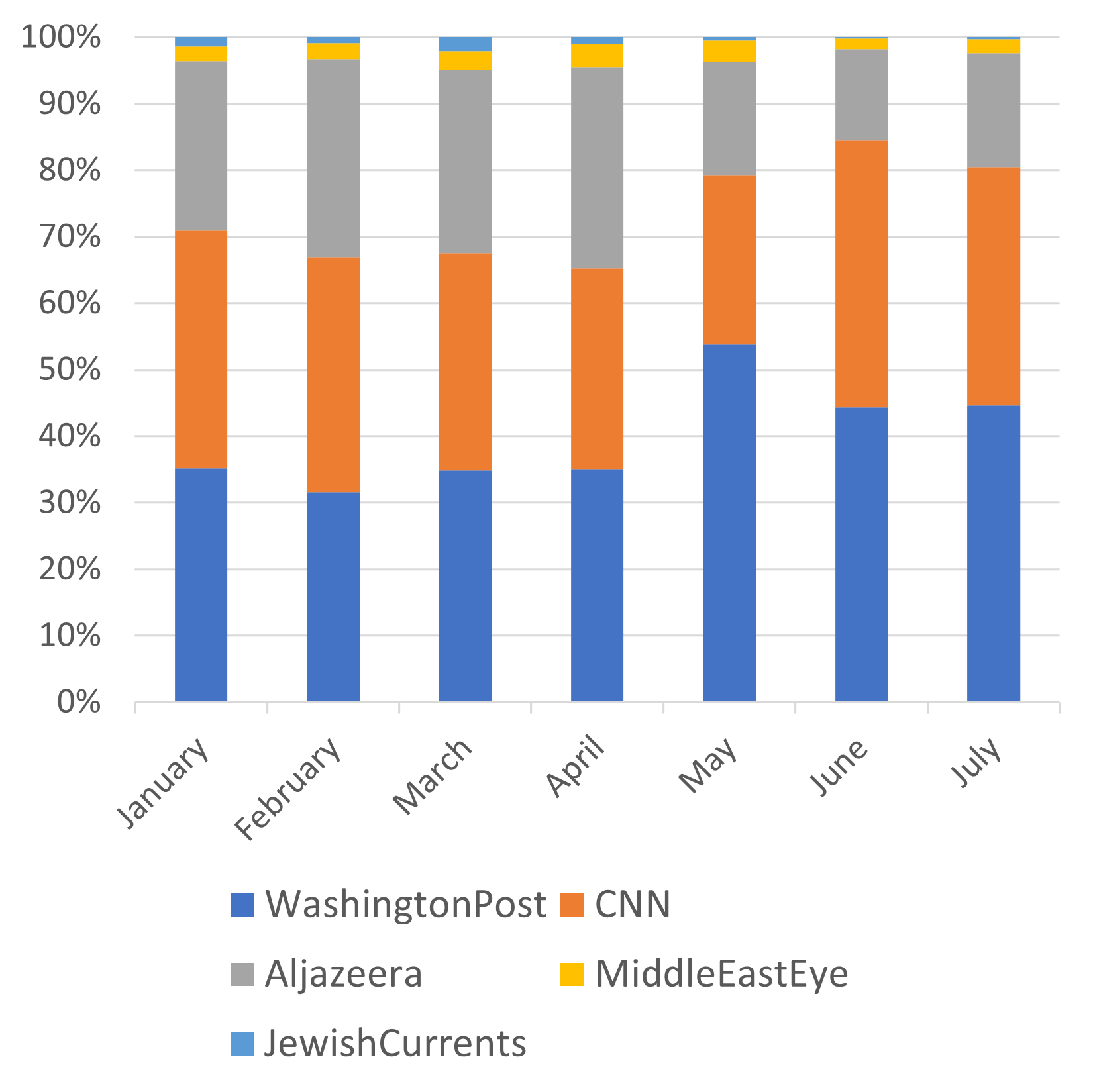}(b)
    \includegraphics[width=.4\linewidth]{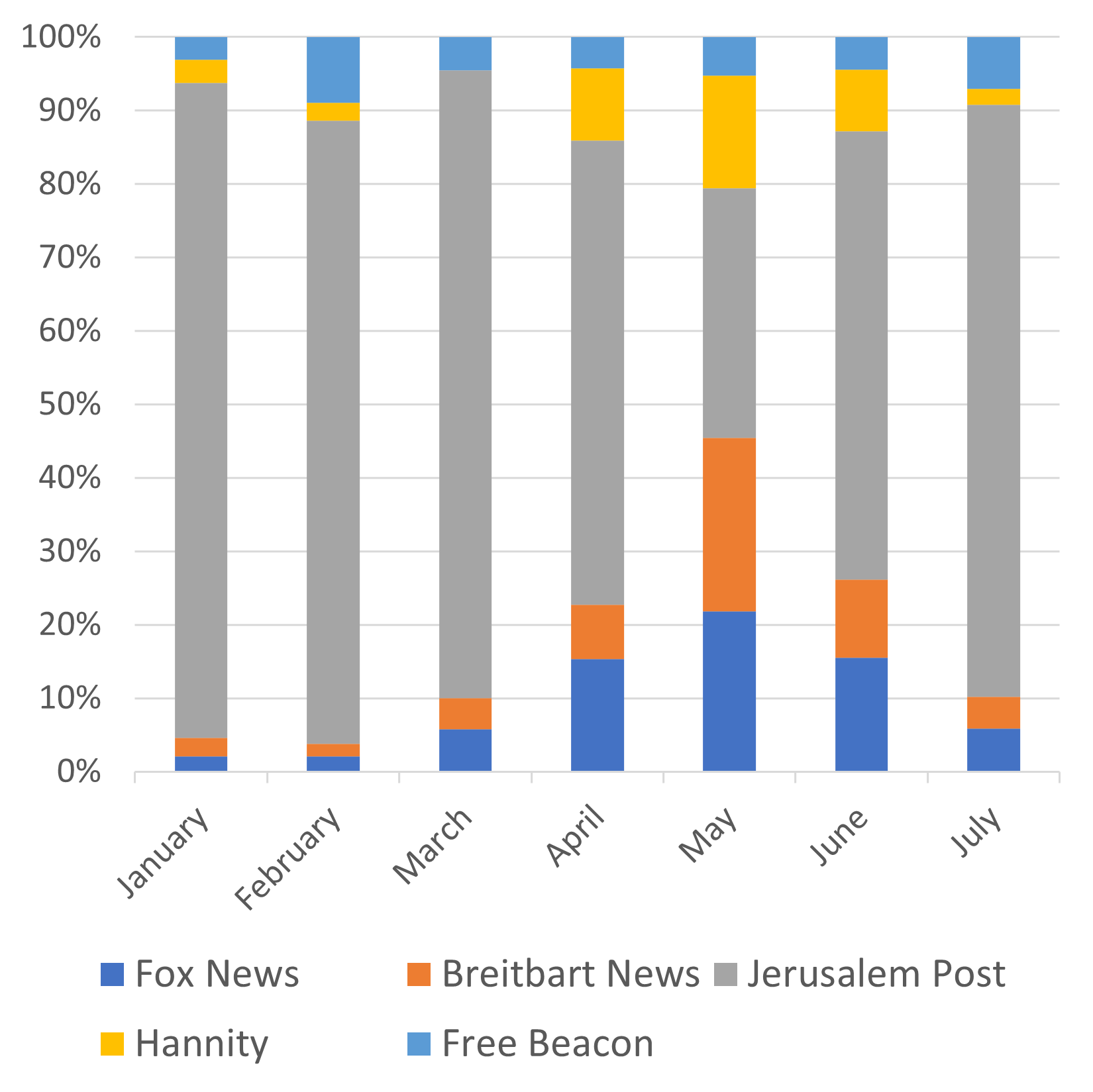}(c)
    \includegraphics[width=.4\linewidth]{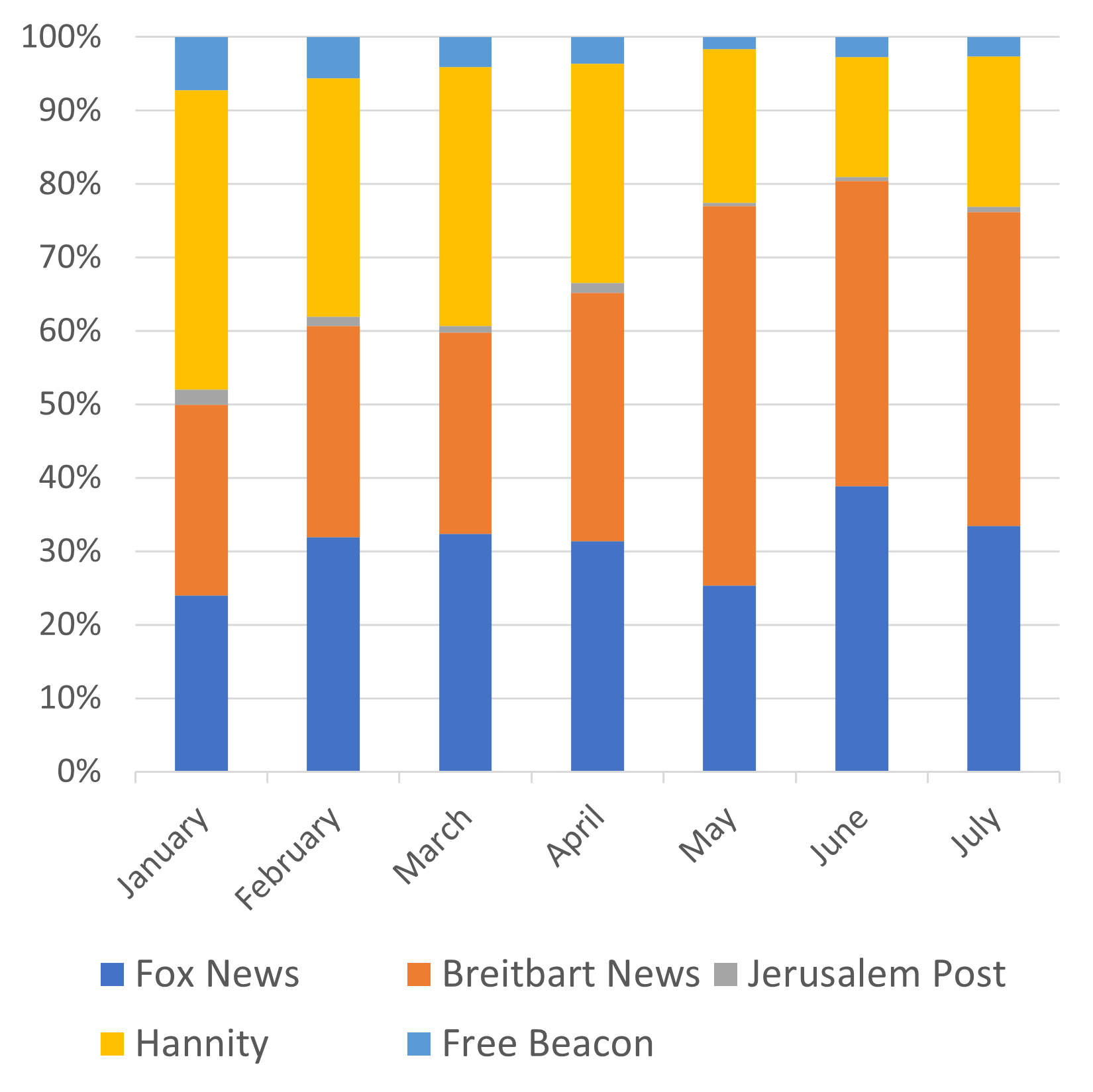}(d)
    \caption{Proportional citation per source for pro-Palestinian group for (a) topic-related and (b) non-topic-related tweets, and for pro-Israeli group for (c) topic-related and (d) non-topic-related tweets.}
    \label{fig:proportionalCitations}
\end{figure*}

As can be seen, both CNN and Washington Post received much criticism (72.3\% \& 21.0\% respectively) and very little praise (5.0\% \& 5.5\% respectively) from the pro-Palestinian group. Unlike CNN, Washington Post articles were often used a source of news.  Further, Washington Post articles were far more likely to be critical of Israel compared to CNN, and such criticism may have contributed to significantly more topically-relevant citations of the Washington Post compared to CNN (Figure \ref{fig:proportionalCitations} (a)).  One curious aspect about CNN was that other journalists and activists reported attacks on CNN staff at the hands of Israeli soldiers.  This hints to CNN having reporters on the ground. Yet, their coverage is criticized by a demographic that typical consumes their content. 

Similarly we analyzed the 3 foreign and less popular sources, and we found 3 main themes in all of them, namely:
\begin{itemize}
    \item Exposing Israeli atrocities:
    \begin{itemize}
        \item Aljazeera (74.9\% -- mostly in the form of news). Ex. RT @AJPlus: Israeli police stormed Jerusalem's Al-Aqsa Mosque hours after the Israel-Hamas ceasefire, firing rubber bullets and stun grenades at Palestinians.  Witnesses say some Palestinians stayed after Friday prayers to celebrate the ceasefire. Police claim there were "riots." \url{https://t.co/Mb8GWgT5OJ}
        \item Middle East Eye (100\%). Ex. RT @MiddelEastEye: I don't know what to do.  A 10-year-old Palestinian girl breaks down while talking to MEE after Israeli air strikes destroyed her neighbour's house, killing 8 children and 2 women  \#Gaza \#Palestine \#Israel \url{https://t.co/PWXsS032F5}
        \item Jewish Currents (13.4\%). Ex. RT @ArielleLAngel: Today in @jewishcurrents, a bit of our own teshuva for our magazine's abandonment of the Palestinian people during the Nakba. Grateful to Dorothy Zellner, a daughter of the Jewish Communist left, for her work on bringing this little-known history to light \url{https://t.co/09WuEVYMkn} 
    \end{itemize}
    \item Criticizing US policy:
    \begin{itemize}
        \item AlJazeera (25.1\%). Ex. RT @AJPlus: The U.S. blocked the UN Security Council from issuing a statement aimed at reducing tensions between Israel and Palestinians, diplomats say.  Sources told @AFP that 14 out of 15 Sec. Council members were in favor of the statement.  The U.S. also blocked a similar text on Monday. \url{https://t.co/xsvkwLTXbr} 
        \item Jewish Currents (39.9\%). Ex. RT @theIMEU: BREAKING: @AOC plans to introduce a bill to block Biden's weapons sale to Israel.  Read more here: \url{https://t.co/1jAVlsUVxF} \url{https://t.co/NaH7yVSTF1} 
    \end{itemize}
    \item Spreading awareness about Palestinian/Israeli conflict:
    \begin{itemize}
        \item Jewish Currents (46.6\%). Ex. RT @JewishCurrents: We’ve put together a selection of our articles from the past few years that provide essential background for the crisis in Israel/Palestine. Here’s a thread of some of them: https://t.co/pWRvniT4xl 
    \end{itemize}
\end{itemize}

\begin{figure}[ht]
    \centering
    \includegraphics[width=0.86\linewidth]{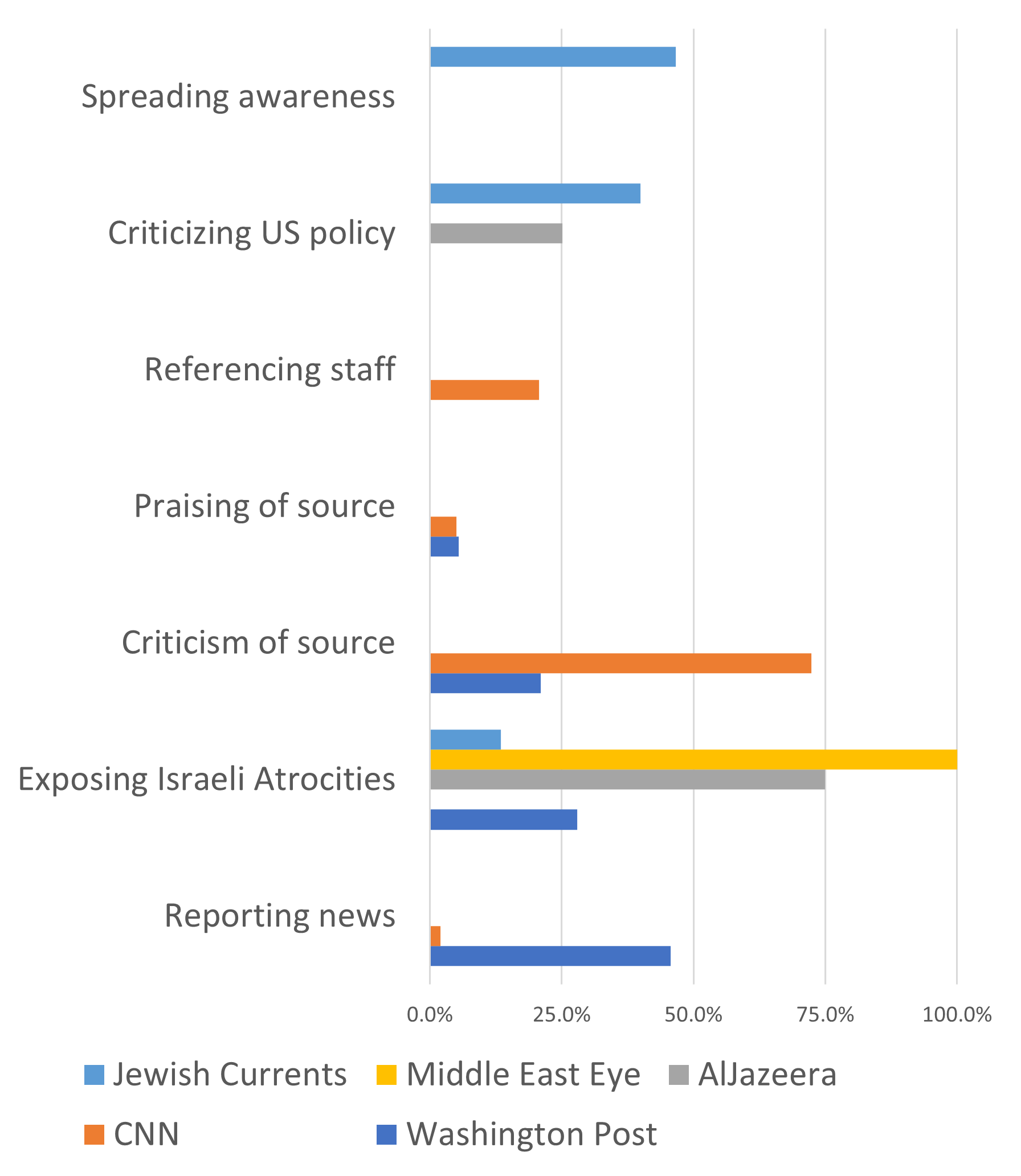}
    \caption{Tweets categories for pro-Palestinian group}
    \label{fig:categoriesProPal}
\end{figure}

Figure \ref{fig:categoriesProPal} summarizes the categories for pro-Palestinian group.  As seen from the references to foreign and less popular sources, the sources report on the conflict from different angles.  For example, AlJazeera focused mostly on providing specific news items (storming of Al-Aqsa compound, number of Palestinian casualties, etc.) that portray the Israeli side in a negative light. Middle East Eye tweets were more sensationally critical of Israel with, for example, an interview with 10 year old girl featuring prominently in their coverage.  Jewish Currents tweets were more concerned with bringing awareness to the plight of the Palestinians and criticizing the US position.

In contrasting all the sources for the pro-Palestinian groups, the answer to the second questions of why users are resorting to foreign or less popular media seems to be a mixture of \textbf{diverging stances from popular sources} such as CNN, \textbf{a desire to learn more} specific news about the conflict (ex. from AlJazeera), and the \textbf{sharing of more sensational content} (ex. from Middle East Eye). Figure \ref{fig:proportionalCitations} (a) shows the relative proportion of topically related citations from the five different sources between January to July 2021. The Figure suggests that foreign and less popular sources dominated a greater proportion of user interest during the conflict period (May, 2021) and in the following months. This reinforces the narrative that users are turning to less popular sources that satisfy their needs.  

\begin{figure}[ht]
    \centering
    \includegraphics[width=0.86\linewidth]{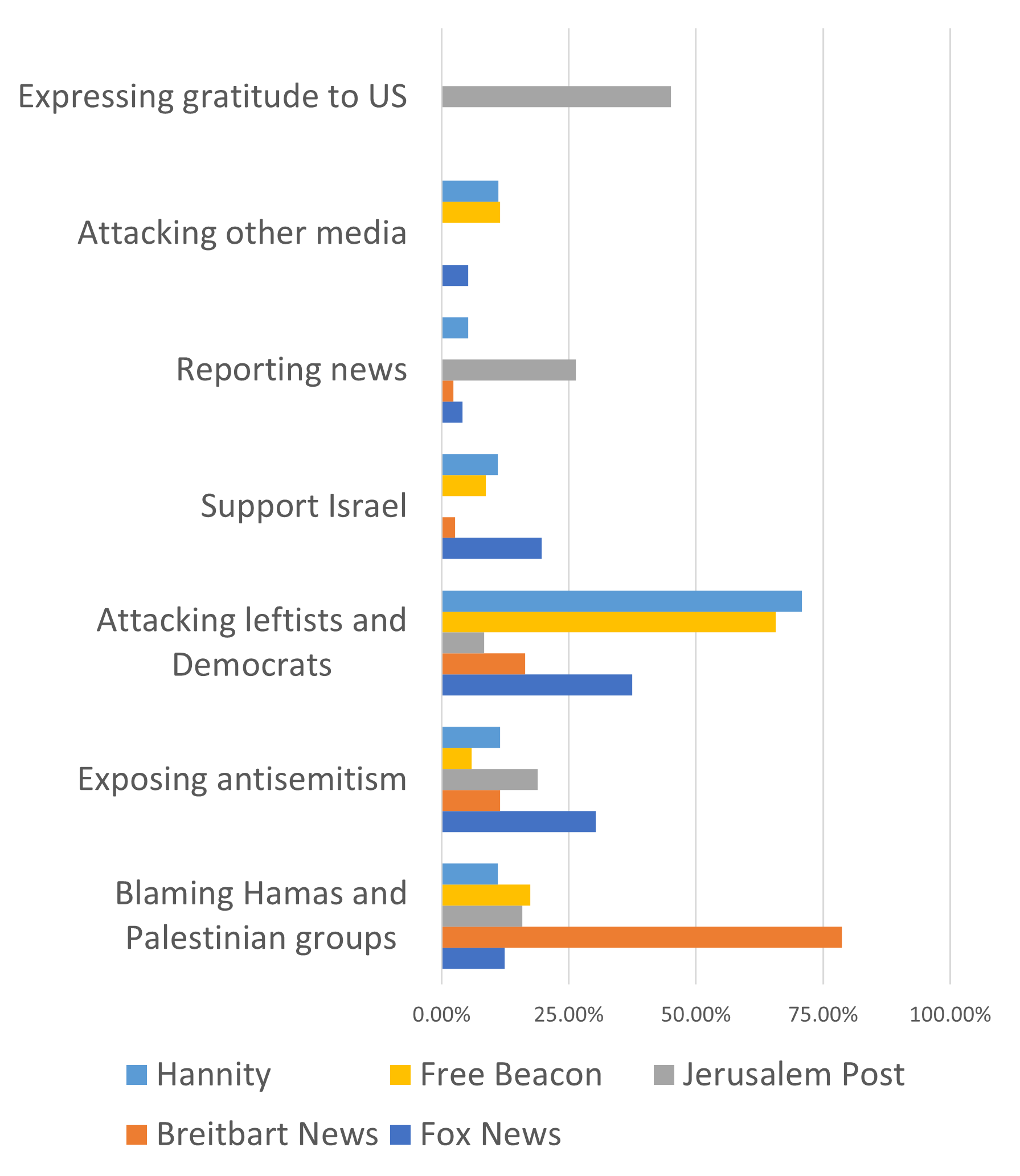}
    \caption{Tweets categories for pro-Israeli group}
    \label{fig:categoriesAntiPal}
\end{figure}

\paragraph{Pro-Israeli Group}
For the pro-Israeli group, we conducted a similar analysis.  For the ``popular'' sources, namely Fox News and Breitbart News.  For Fox News, the categories are:

\begin{itemize}
    \item Attacking leftists and Democrats including Biden, Black Lives Matter (BLM), and the ``Squad'' (group of progressive lawmakers) (37.5\%). Ex. RT @RichardGrenell: This makes me sick. It’s disgusting and every American should condemn this immediately.   \url{https://t.co/I8WavM578M} -- (citing a Fox News article about BLM solidarity with Palestinians).
    \item Exposing antisemitism (30.3\%).  Ex. @DreyfusShawn: Why is Fox the only major network covering the antisemitic attacks on Jews by Palestinians in the US? 
    \item Supporting Israel (19.7\%). Ex. RT @Mike\_Pence: If the World Knows Nothing else Let the World Know this-  America Stands With Israel \url{https://t.co/i0waMKe2al} 
    \item Blaming Hamas and Palestinian groups (12.4\%). Ex. RT @TomCottonAR: There will be a `significant de-escalation' when Hamas stops its terror attacks against Israel. \url{https://t.co/pdftQttXFg}
    \item Attacking other media sources (5.2\%). Ex. RT @BoSnerdley: CNN avoids Israel-Hamas conflict during primetime, spends over 90 minutes on Liz Cheney \url{https://t.co/0lmHznAK98} \#FoxNews 
\end{itemize}

As for Breitbart News, the categories are:

\begin{itemize}
    \item Blaming Hamas and Palestinian groups (78.7\%). Ex. RT @BreitbartNews: Gaza has spent \$175 million on terror rockets in a week but has no money for vaccines or education. Let that sink in for a moment.
    \item Attack leftists and Democrats including Biden, Black Lives Matter, the ``Squad'' (16.4\%). Ex. RT @BreitbartNews: Andrew Yang did the right thing and backed Israel. Now Democrats are making him pay the price. \url{https://t.co/Y2C8n5q9Yo} 
    \item Exposing anti-Semitism (11.5\%). Ex. RT @BreitbartNews: Pro-Palestinian demonstrators were filmed Thursday as they attacked Jews in Manhattan. \url{https://t.co/0JEvY8zdFK} 
    \item Supporting Israel (2.7\%). Ex. RT @BreitbartNews: Christians from around the globe have donated nine portable bomb shelters to Israeli communities near the Gaza border. \url{https://t.co/Nda4F9eoiH} 
\end{itemize}

For the foreign/less popular sources, Sean Hannity and Free Beacon were very similar in their framing and very different from the Jersualem Post.  For the Jerusalem Post, the main categories are:

\begin{itemize}
    \item Expressing gratitude to US, including Biden and Democrats (45.1\%). Ex. RT @Jerusalem\_Post: House Speaker \#NancyPelosi condemned the ``escalating and indiscriminate rocket attacks by \#Hamas against \#Israel'' who ``has the right to defend herself against this assault, which is designed to sow terror and undermine prospects for peace.''
    \item Reporting news (26.4\%).  Ex. RT @allahpundit: Israel showed US ‘smoking gun’ on Hamas in AP office tower, officials say \url{https://t.co/8sC8Ri3SG1} 
    \item Exposing antisemitism (18.8\%). Ex. RT @Jerusalem\_Post: Following a spate of anti-Israel protests across Germany tied to the ongoing \#Israel-\#Gaza violence, political leaders here have vowed to crack down on demonstrators who have used \#antisemitic rhetoric and have attacked Jewish institutions.
    \item Blaming Hamas and Palestinian groups (15.9\%). Ex. RT @Jerusalem\_Post: \#BREAKING: New Illustrated images show exactly how \#Hamas operate from within civilian neighborhoods in the \#Gaza Strip. Read more: \url{https://bit.ly/3u4SyWe} 
    \item Warning talks with Iran (8.3\%). RT @Cliff\_Sims: JohnRatcliffe in @Jerusalem\_Post: ``If the Biden administration continues on its current course, re-enters a nuclear deal with Iran and lifts sanctions on the regime, this will be tantamount to giving them another airplane full of cash.'' \url{https://t.co/hNEFOJQjow}  
\end{itemize}

For Sean Hannity and Free Beacon the main categories are:

\begin{itemize}
    \item Attacking leftists and Democrats including Biden, BLM, and the Squad:
    \begin{itemize}
        \item Sean Hannity (80.7\%). Ex. RT @seanhannity: TRUMP on ISRAEL: ‘Under Biden the World is Getting More Violent and More Unstable’ \url{https://t.co/ltRpYebxgD} 
        \item Free Beacon (65.7\%). Ex. RT @EliLake: Read @continetti on the corbynization of the Democratic party. \url{https://t.co/ishI4YnPPO} 
    \end{itemize}
    \item Blaming Hamas and Palestinian groups:
    \begin{itemize}
        \item Sean Hannity (11.0\%). Ex. RT @seanhannity: Report: Iran Funding Palestinian Terrorist Attacks On Israel \url{https://t.co/RlV5HDI9f0} 
        \item Free Beacon (17.4\%). Ex. RT @Kredo0: FLASHBACK to March: Palestinians Funneled Hundreds of Millions to Terrorists, State Dept Report Reveals \url{https://t.co/hiThij4hj9} 
    \end{itemize}
    \item Supporting Israel:
    \begin{itemize}
        \item Sean Hannity (11.0\%). Ex. RT @seanhannity: CRUZ to VISIT ISRAEL: The Senator Will Travel to Israel in the ‘Coming Days’ to Assess Security Situation \url{https://t.co/hRjmn7xAlX} 
        \item Free Beacon (8.7\%). Ex. RT @SenTedCruz: “Sens. Cruz and Hagerty Land in Israel to Assess Damage from Hamas War” via @Kredo0 @FreeBeacon \url{https://t.co/LKkfOgsSgI} 
    \end{itemize}
    \item Exposing anti-Semitism:
    \begin{itemize}
        \item Sean Hannity (11.5\%). Ex. RT @seanhannity: HATE IN LA: Pro-Hamas Activists Attack Jewish Americans at a Restaurant in Los Angeles \url{https://t.co/FIKVh5GVh4} 
        \item Free Beacon (5.9\%). Ex. RT @MatthewFoldi: new from @alexnester2020 and me @FreeBeacon  @NorthwesternU profs @NUQatar are quick to condemn Israel  I reached out to 27 who signed an anti-Israel statement asking if their avowed support for Asians extends to Asian slaves in Qatar. No response  \url{https://t.co/DYBk5X4tSs}  (1/8) 
    \end{itemize}
    \item Attacking other media sources:
    \begin{itemize}
        \item Sean Hannity (11.1\%). Ex. RT @seanhannity: Facebook Shuts Down Prayers for Israel Page with Over 77 Million Followers \url{https://t.co/peuuI3qH9x} 
        \item Free Beacon (11.5\%). Ex. RT @BoSnerdley: AP Hires Anti-Israel Activist as News Associate \url{https://t.co/dVAR2I7ON2} 
    \end{itemize}
\end{itemize}

Figure \ref{fig:categoriesAntiPal} summarizes the categories for the media sources for the pro-Israeli group.  As the data shows, the foreign source (Jerusalem Post) was the most different from all the other sources in that it focused on expressing gratitude to the US administration, including Biden and the Democrats, and reported on events on the ground.  The less popular US sources (Hannity and Free Beacon) are highly partisan with the majority of the tweets associated with attacks on Democrats, liberals, and progressives. Breitbart News was mostly concerned with assigning blame to Palestinians. Fox News covered the different categories more evenly.  

In contrasting all the sources for the pro-Israeli group, the answer to the second questions of why users are resorting foreign or less popular media seems to be a mixture of \textbf{more extreme partisanship or sensationalism} and a \textbf{desire to learn more specific news} about the conflict.  Both these aspects are shared with the pro-Palestinian group.  However, unlike the case of the pro-Palestinian group, the stance of the popular media sources was not divergent from the foreign and less popular ones, and the popular media sources were not eclipsed by the foreign and less popular ones.  Figure \ref{fig:proportionalCitations} (c) shows the relative proportion of topically related citations from the five different sources between January to July 2021. The Figure suggests that Jerusalem Post was the main source of topical information before the conflict.  With the advent of the conflict, it was crowded by popular sources.  However, as the conflict subsided, Jerusalem Post came back to reclaim a dominant position.  This is the exact opposite of what it is observed for the pro-Palestinian group.  We postulate that when the narrative of the foreign or less popular sources contradicts that of the popular sources, such in the case of pro-Palestinian group, such sources would dominate in the time of conflict.  However, when such sources are in line with the popular sources, the popular sources gain relative market share in the time of conflict.

The third question of interest is: \textbf{Does the sudden boost in popularity of the sources translate into future popularity?} Figures \ref{fig:proportionalCitations} (b) and (d) show proportional user interactions for the non-topically related tweets for the pro-Palestinian and pro-Israeli groups respectively.  As the data suggests, while foreign and less popular sources benefited from their topically-related coverage during and after the conflict, the benefit did not translate into non-topically related coverage.  In fact, their relative share for non-topically related tweets decreased after the conflict compared to before the conflict. This seems to be the case for all foreign and less popular sources for both groups.  It is unclear whether this is transient or more permanent, and more data from later time epochs are required to reach a more definitive conclusion.  Nonetheless, the answer to the third question seems to be that the sudden boost in popularity may not translate into future popularity. We suspect two potential reasons for this.  First, previous research suggests that well-established media organizations are resilient in the face of nascent media organization \citep{nelson2020enduring}.  Second, the event may actually frame a media source, where it becomes strongly associated with specific topics.  Further work is required to confirm our suspicion.

We looked at the tweets of AlJazeera and Sean Hannity for the pro-Palestinian and pro-Israeli groups respectively, as both maintained a sizeable, though decreased, proportion of user interest after the conflict.  From inspecting the tweets related to AlJazeera, we do not find clear pattern that would definitively explain the post-conflict trend.  We found among the top 50 retweeted tweets: 24 are related to the US (retweeted 565 times); 24 are related to foreign news (retweeted 530); and the 2 remaining ones are related to a global issue, namely climate change (retweeted 91 times).  We also notice that most of AlJazeera retweets were from @AJPlus, which is a social media arm of Aljazeera -- @AJPlus retweeted 3,483 times compared to 2,194 retweets of @AJEnglish (Aljazeera English).  As for the 50 most retweeted tweets associated with Sean Hannity: 30 tweets are attacking Biden and the democrats (retweeted 862 times), 13 tweets are praising Trump (retweeted 416 times), and the remaining are a mix of different news and political issues.  Such a distribution highlights the politically sensational nature of Sean Hannity.

\section{Discussion and Conclusion}
In this paper, we examined the interaction of users with foreign and less popular media sources in conjunction with a major polarizing event, specifically the May 2021 Israeli-Palestinian conflict.  We conducted our analysis on more than 8 thousand users, whom we automatically labeled as pro-Palestinian or pro-Israeli.  We show that users may resort to consuming and/or referring to less popular or foreign sources during a polarizing event.  We identified multiple reasons for this behavior.  First is to find content that reaffirms their worldview, specially when such view is not being expressed by popular media sources.  This was clear for the pro-Palestinian group, where they were sharing content from AlJazeera and Middle East Eye that was mostly critical of Israel, and they were critical of the coverage of popular left-leaning sources such as CNN and the Washington Post. Hence, news consumers may sideline popular sources that they typically consume in favor of other source for specific topics.  Second, users may be interested in more extreme, hyper-partisan, or sensational content, such as Sean Hannity for the pro-Israeli group and Middle East Eye for the pro-Palestinian group.  Third, users may have a desire to learn more about a topic, particularly when more popular sources do not provide sufficient coverage or they provide more opinionated content.  We observed this in the case of Jerusalem Post and AlJazeera, both of whom were reporting on events on the ground, though they may have been framed in ways that reaffirm specific viewpoints.

Less popular and foreign sources may challenge the dominance of popular sources on specific topics if a substantial portion of the typical consumers of popular sources are unhappy with their narrative -- as in the case of the pro-Palestinian group. If the stance of the foreign/less popular sources is aligned with the stance of the popular sources, then popular sources may actually crowd out less popular sources that typically dominate a topic in normal times. 

Given the increased popularity of foreign or less popular source on a specific topic may not translate to popularity on other topics either during or after a major event. In fact, we observed that relative percentage of citations of sources such as AlJazeera and Hannity for non-topical tweets actually decreased from May onward.  Given the time horizon, we cannot ascertain if this decrease is transient or sustained.  However, we can clearly observe no increase.  Conversely, popular sources, such as CNN and Fox News, may experience some drop in relative popularity among portions of their typical consumers during the time of the major event.  However, they seem to rebound after the event has passed.  Further investigation using other events and longer time horizons are needed to confirm this finding. Another interesting note, particularly concerning CNN, is that pro-Palestinian users were in fact consuming CNN content in general (see Figure \ref{fig:proportionalCitations} (b)), but they were choosing not share conflict related content from the site (Figure \ref{fig:proportionalCitations} (a)).

\bibliographystyle{aaai}
\bibliography{references}

\end{document}